\documentclass[10pt]{article}

\usepackage{amsmath}
\usepackage{amssymb}

\usepackage{graphicx}

\usepackage{cite}

\usepackage{color} 

\usepackage[labelfont=bf,labelsep=period,justification=raggedright]{caption}

\makeatletter
\renewcommand{\@biblabel}[1]{\quad#1.}
\makeatother

\date{}

\begin{document}

\begin{flushleft}
{\Large
\textbf{Spatio-temporal variation of conversational utterances on Twitter}
}
\\
Christian M. Alis, 
May T. Lim$^{\ast}$
\\
National Institute of Physics, University of the Philippines, Diliman, 1101 Quezon City, Philippines
\\
$\ast$ E-mail: mlim@nip.upd.edu.ph
\end{flushleft}

\section*{Abstract}

Conversations reflect the existing norms of a language. Previously, we found that utterance lengths in English fictional conversations in books and movies have shortened over a period of 200 years. In this work, we show that this shortening occurs even for a brief period of 3 years (September 2009-December 2012) using 229 million utterances from Twitter. Furthermore, the subset of geographically-tagged tweets from the United States show an inverse proportion between utterance lengths and the state-level percentage of the Black population. We argue that shortening of utterances can be explained by the increasing usage of jargon including coined words.


\section*{Introduction}

Utterances, the speaking turns in a conversation, relay short bits of information.  Though utterances adapt strongly to medium \cite{alis_adaptation_2012}, utterance shortening has been observed over a span of two centuries. Here we show that utterances in the online social medium Twitter did not only significantly shorten in a span of a few years but also varied geographically---providing evidence of increasing usage of jargon brought about by formation of groups. Our use of Twitter conversations provided us with a large, highly resolved and current dataset.

Twitter (\texttt{twitter.com}) is an online social medium that allows its users to post messages (\textit{tweets}) of up to 140 characters in length, which are public by default. Previous studies~\cite{ritter_unsupervised_2010,kumar_dynamics_2010} on Twitter conversations focused on modelling the structure of conversations rather than the form of utterances. Recent studies have ranged from characterizing the graph of the Twitter social network ~\cite{huberman_social_2008, kwak_what_2010} to inferring the mood of the population~\cite{goncalves_modeling_2011,bollen_twitter_2011,golder_diurnal_2011,kloumann_positivity_2012}. Owing to the large number of Twitter users (about half-billion in June 2012~\cite{semiocast_twitter_2012}) and easy access via the provided application programming interfaces (API), Twitter has become a platform for studying the usage of the English language. For example, it has been found that longer Twitter messages (\textit{tweets}) are more likely to be credible~\cite{odonovan_credibility_2012} and, by determining where tweets were posted, dialects~\cite{eisenstein_latent_2010} and geographical diffusion of new words~\cite{eisenstein_mapping_2012} are observable.

Conversations in Twitter are typically performed in one of two ways: privately, using \textit{direct messages}; or publicly, using \textit{replies}. Replies~\cite{_what_2012} are tweets that begin with the username of the recipient prefixed with an at (@) sign, for example, \textit{@bob Hello! How are you?}. Since replies may be viewed by other users aside from the recipient, replies are used for public conversations~\cite{williams_how_2008} akin to having conversations while other people are listening.

Conversation analysis usually investigates the structure of conversations~\cite{wooffitt_conversation_2005} by looking at the interaction of utterances instead of the individual utterances themselves. Since we are more interested with the encoding of information or idea into an utterance, this paper focused instead on the construction of individual utterances and not in their interaction. More specifically, the length of utterances are measured because the production time and the amount of information of an utterance should be correlated with its length.

Sentence lengths are not as widely studied as words, and conversational utterances less so. The study of sentence lengths began with the work of Udny Yule~\cite{yule_sentence-length_1939} in 1939 and eventually led to the discovery that sentence length distributions may be approximated by a gamma distribution~\cite{sigurd_word_2004}. On the other hand, the mean length of utterance is used to evaluate the level of language development of children~\cite{klee_relation_1985,dollaghan_maternal_1999}. The length of sentences and utterances are usually measured in terms of words or morphemes but we used the number of characters (\textit{orthographic length}) as unit of length because Twitter imposes a maximum tweet length in terms of characters. The use of orthographic length of sentences has previously been shown to be a valid unit when comparing utterance length distributions~\cite{alis_adaptation_2012}. Furthermore, the orthographic length of words is highly correlated with word length in terms of syllables~\cite{strauss_word_2006}.

In this paper, 229 million conversational utterances collected from 18 September 2009 to 14 December 2012 are first characterized. By comparing expected (fitted) and empirical utterance length distributions, we show that the character limit of tweets has little to no effect on the median utterance length. We also identify some factors that significantly affect the utterance length. The dataset is then disaggregated to reveal that utterances shortened in a span of more than three years. Possible mechanisms of shortening are then explored. Finally, the variation of utterance length across different US states and  its correlation with demographic and socioeconomic variables are investigated.

\section*{Results}

\subsection*{Aggregate utterance length distribution}

\begin{figure}[!ht]
\begin{center}
\includegraphics[width=4in]{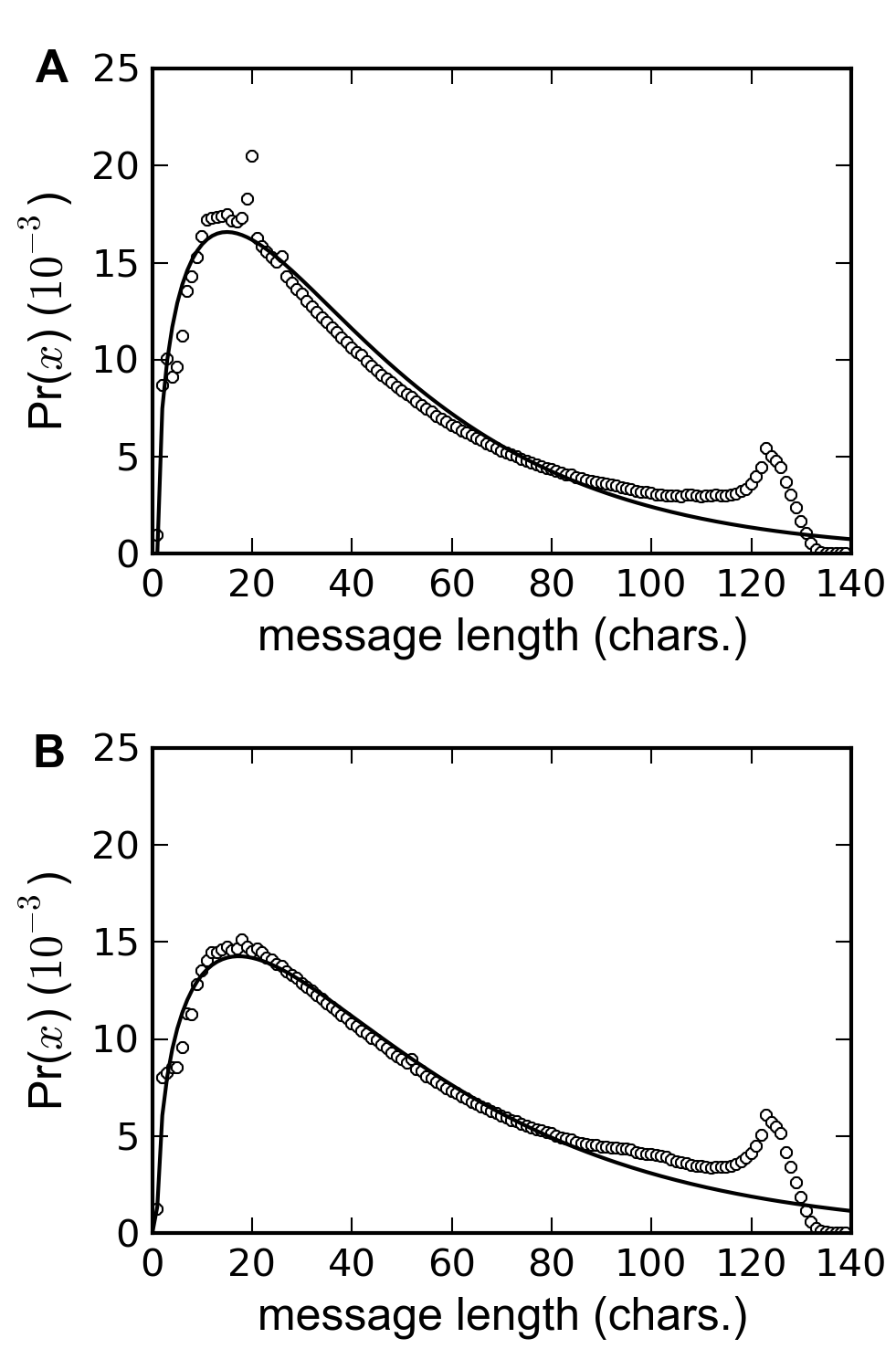}
\end{center}
\caption{
{\bf Utterance length distribution of the entire dataset.} \textbf{A.} Unfiltered utterance length distribution of the entire dataset \textbf{B.} Utterance length distribution of English tweets with URLs and LOL removed. The solid line in both plots is the best fit of Eq.~\eqref{eq:gammafit}.
}
\label{fig:agg}
\end{figure}

The utterance length distribution (ULD) of the entire data set (Fig.~\ref{fig:agg}A) is bimodal and can be fitted with a gamma distribution after taking the 140-character limit into account \cite{alis_adaptation_2012}. It is bimodal due to the mixture of the natural (unconstrained) ULD and shortened (constrained) ULD forced by the 140-character limit. To estimate the unconstrained ULD, a generalized gamma distribution,
\begin{equation}
\label{eq:gammafit}
\mathrm{Pr}(x) = \frac{\tilde{x}^{\alpha - 1} e^{-\tilde{x}}}{{\Gamma(\alpha)}},
\end{equation}
where ̃$x$ is the utterance length, $\tilde{x} = (x-x_0)/s$ is the scaled utterance length, and $\alpha$, $x_0$ and $s$ are fitting parameters that describe the shape, translation and ordinate scaling factor, respectively, was fitted on the utterance length distribution from $x=1$ char. to a cut-off length $x=x_c$ using least squares as was done in Ref. \cite{alis_adaptation_2012}. The estimated natural ULD ($\alpha=1.46$, $x_0=1.01$ char., $s=30.0$ char.) fits the empirical ULD  with an $r^2=0.950$.

Both empirical and fitted (unconstrained) ULD are skewed to the right and the quartiles (Q1=25th percentile; Q2=median=50th percentile; Q3=75th percentile) are either the same (Q1=19 char., Q2=36 char.) or differs by 3 characters (Q3$_\text{empirical}=65$ char., Q3$_\text{fitted}=62$ char.). From here on, we used the quartiles of the empirical ULD to describe the distributions.

Starting 10 October 2011, all URLs in tweets are automatically shortened by Twitter~\cite{twitter_t.co_2012} into a 20-character URL (\texttt{http://t.co/xxxxxxxx}) and this caused the spike at $x=20$ char. in the ULD. The spike at $x=26$ char. is due to non-English tweets while the spike at $x=3$ char. is due to the acronym LOL (laughing out loud). Restricting utterances to English and removing URLs and LOL result to a smoother ULD (Fig.~\ref{fig:agg}B) but with the same quartiles as the original distribution.

\subsection*{Temporal dependence of utterance lengths}

\begin{figure}[!ht]
\begin{center}
\includegraphics[width=3.27in]{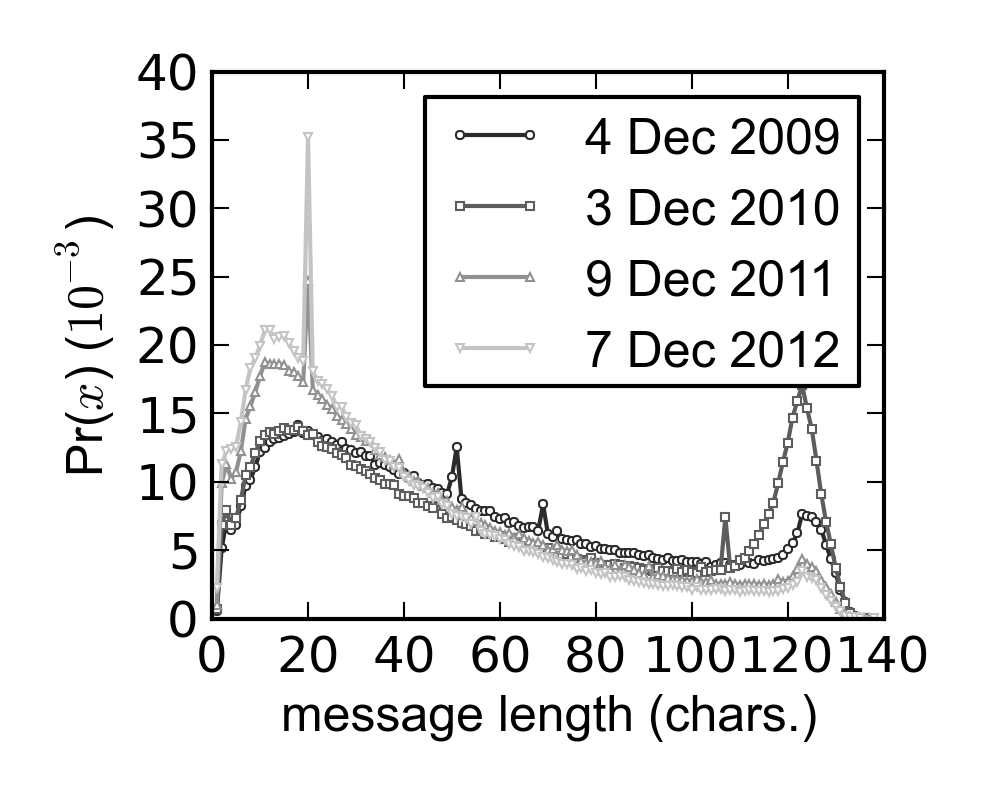}
\end{center}
\caption{
{\bf Representative utterance length distributions per year.} Utterance length distribution of every first available Friday of December in the dataset.
}
\label{fig:strip-utterances}
\end{figure}

\begin{figure}[!ht]
\begin{center}
\includegraphics[width=4in]{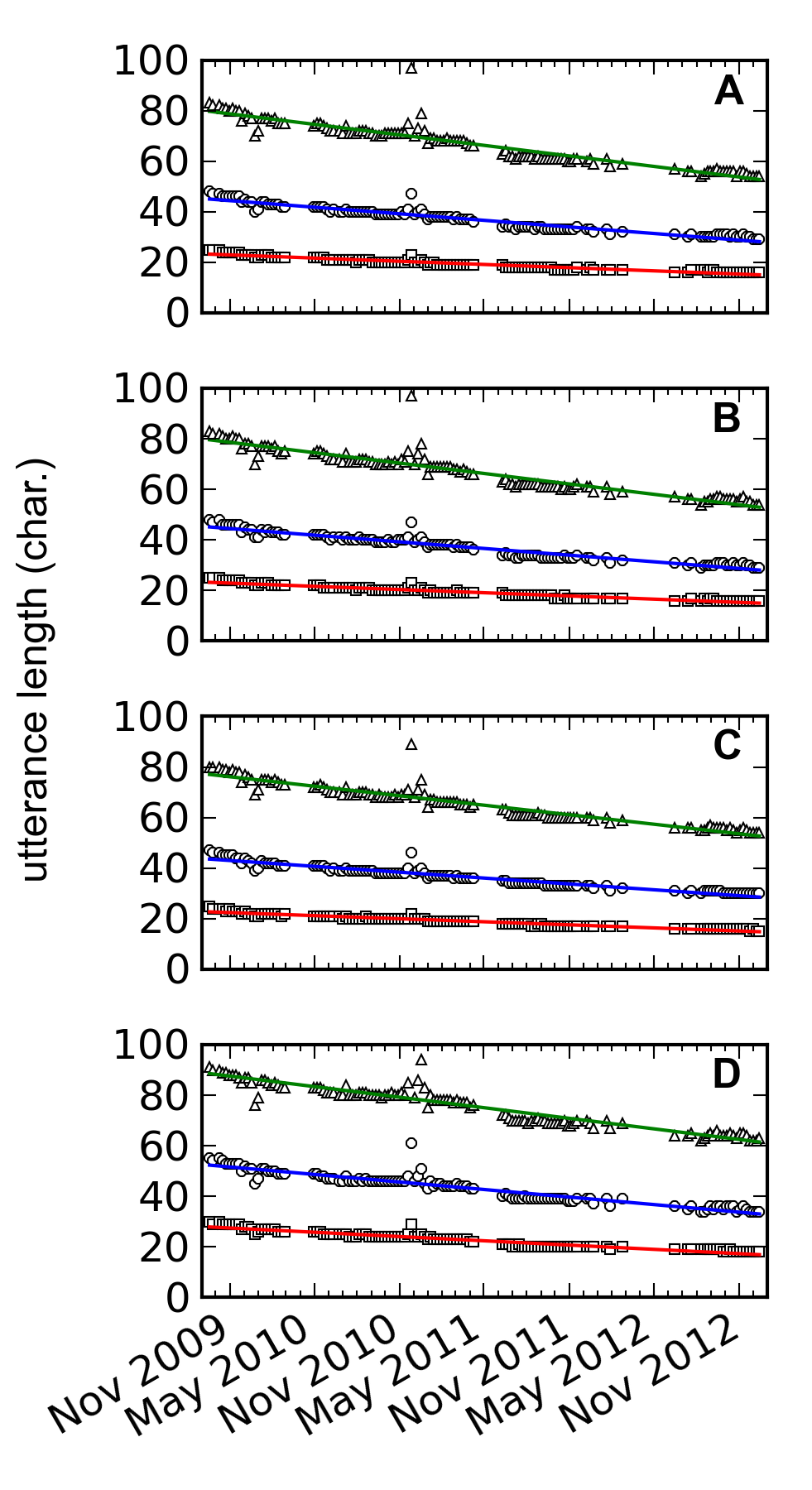}
\end{center}
\caption{
{\bf Utterance length distribution over time.}  First quartile Q1 (square), median Q2 (circle) and third quartile Q3 (triangle) of the \textbf{A.} original dataset, \textbf{B.} after resampling into $10^5$ utterances per day, \textbf{C.} removing URLs and \textbf{D.} restricting to English tweets.
}
\label{fig:temporal}
\end{figure}

Utterance length distributions for tweets aggregated over a 24-hour period that were sampled during Fridays follow the general characteristics of the utterance length distribution for the entire dataset as shown by the representative utterance length distributions in Fig.~\ref{fig:strip-utterances}. The right peak of the plots seems to get smaller and shifted to the left as the date becomes more recent, suggesting shortening of utterances over time. This shortening is clearly shown when the quartiles are plotted with respect to time (Fig.~\ref{fig:temporal}A). The quartiles roughly follow their corresponding regression line except for 26 Nov 2010, which shows an unexpected spike due to spam.

\begin{figure}[!ht]
\begin{center}
\includegraphics[width=4in]{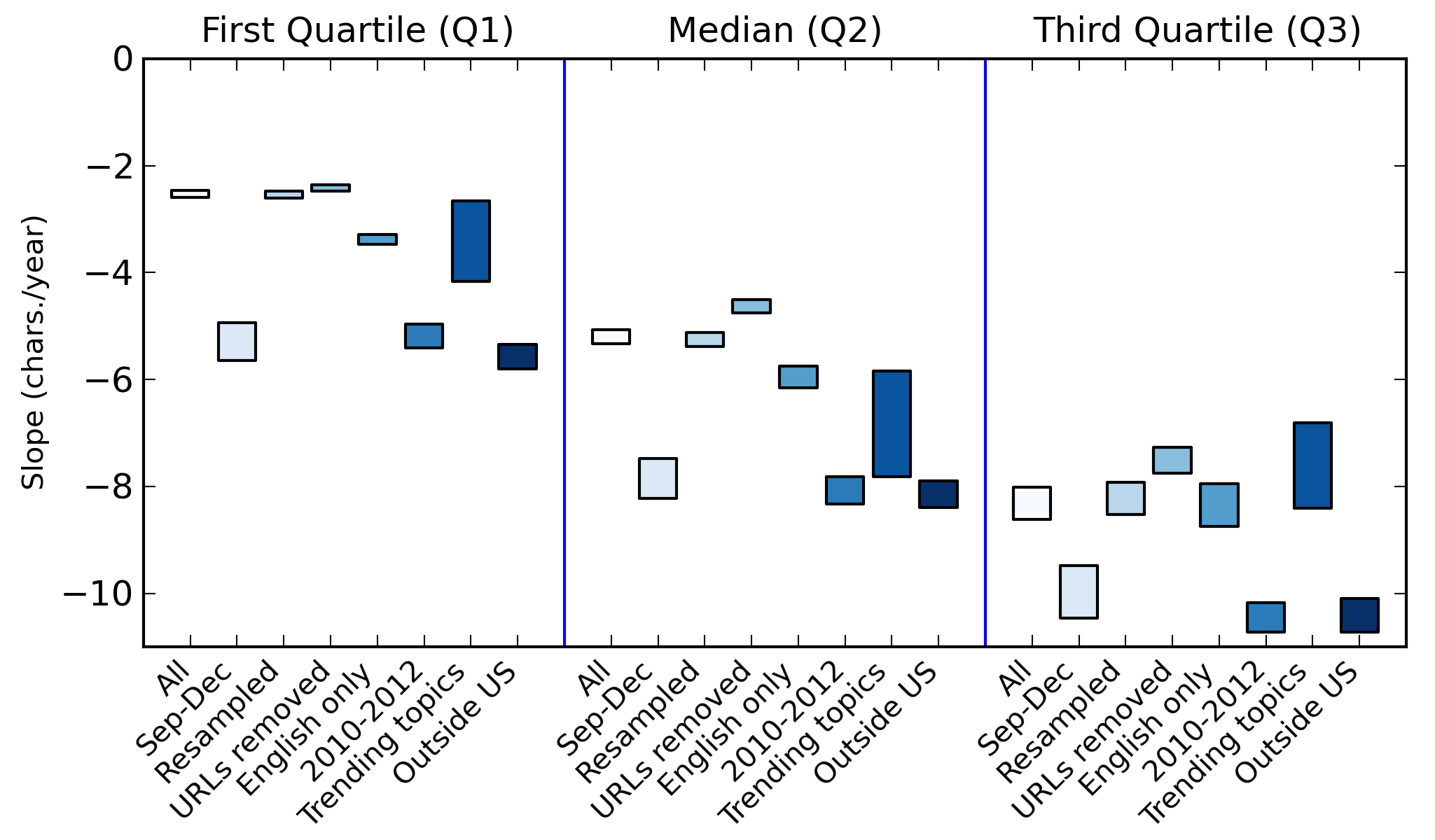}
\end{center}
\caption{
{\bf Slopes of utterance length quartiles temporal regression lines.}  Visualization of Table~\ref{tab:slopes}.
}
\label{fig:temporal-summary}
\end{figure}

As expected the linear regression line of the median (2nd quartile) is not at the middle of Q1 and Q3 regression lines because the utterance length distributions are skewed. The regression line of Q1 (Table~\ref{tab:slopes} and Fig.~\ref{fig:temporal-summary}, all) is less steep than the regression line for the median, which, in turn, is less steep than the regression line for Q3. The shortening of utterances is, therefore, mostly due to the decreased occurrence of longer utterance lengths rather than the shifting of the whole utterance length distributions to the left.

Table~\ref{tab:slopes} and Fig.~\ref{fig:temporal-summary} demonstrate the robustness of the decrease in utterance length. The months included in the dataset differ for each year yet shortening is still observed even if only utterances from the common included months of September to December are considered (Table~\ref{tab:slopes} and Fig.~\ref{fig:temporal-summary}, Sep--Dec). Similarly, the number of utterances per day and the percent of public data collected are not constant throughout the entire dataset. To remove any size effects on the results, $10^5$ utterances, an amount slightly smaller than the smallest daily sample size, were sampled without replacement for each day (Fig.~\ref{fig:temporal}B) but the same observations remained (Table~\ref{tab:slopes} and Fig.~\ref{fig:temporal-summary}, resampled). Another possible reason for the shortening is the increased usage of link shorteners. However, the shortening trend (Table~\ref{tab:slopes} and Fig.~\ref{fig:temporal-summary}, URLs removed) persisted even if all links in the utterances were removed (Fig.~\ref{fig:temporal}C). Finally, restricting the analysis to only English tweets (Fig.~\ref{fig:temporal}D) resulted to the same observations (Table~\ref{tab:slopes} and Fig.~\ref{fig:temporal-summary}, English only).  
 
\subsection*{Possible mechanisms for shortening}

A possible mechanism for utterance length shortening is the shortening of the most frequent words either by a change in orthography (spelling) of the most frequent words or their replacement by shorter words. 

\begin{figure}[!ht]
\begin{center}
\includegraphics{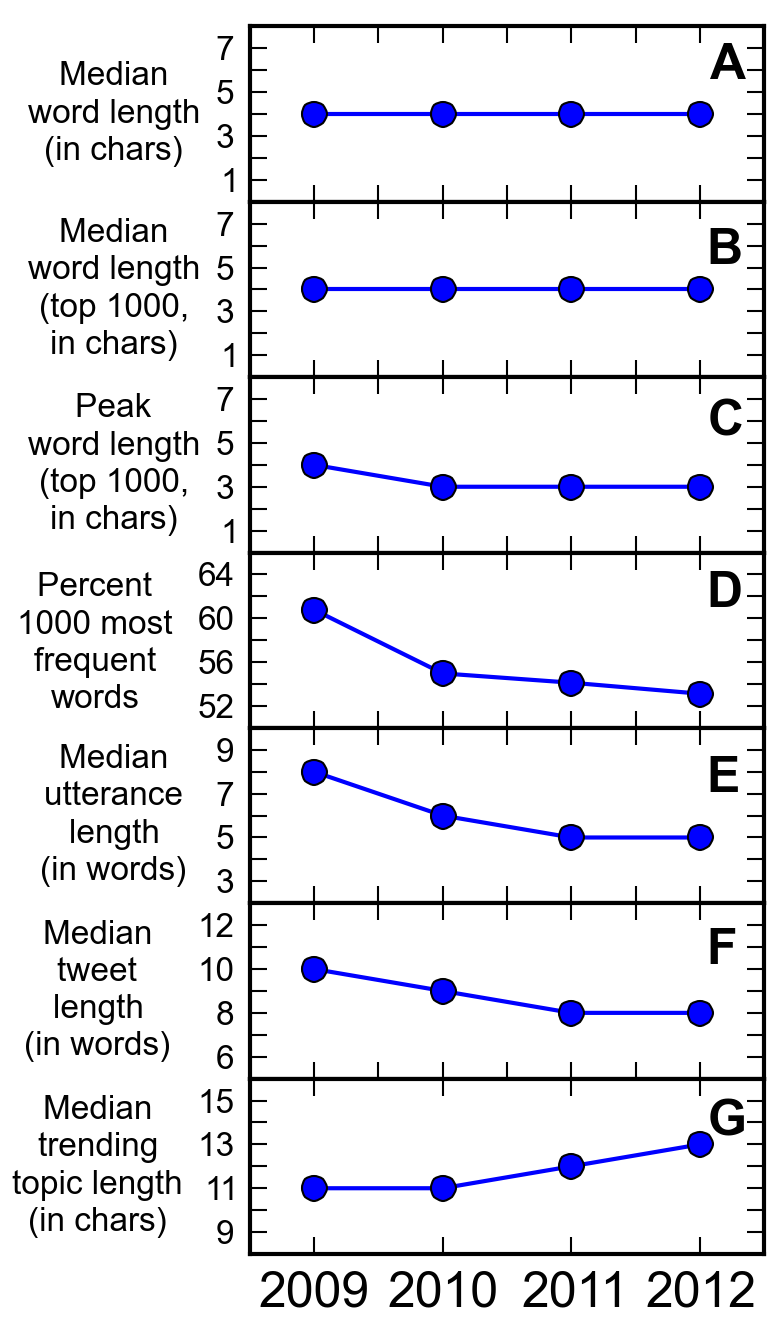}
\end{center}
\caption{
{\bf Exploring possible mechanisms of shortening.} Annual values of \textbf{A.} median word length of all words, \textbf{B.} median word length of the 1000 most frequently occurring words, \textbf{C.} most frequent word length of the 1000 most frequently occurring words, \textbf{D.} fraction of 1000 most frequently occurring words relatively to all occurrences of words, \textbf{E.} median utterance length in number of words \textbf{F.} median tweet length in number of words, and \textbf{G.} median trending topic phrase length.}
\label{fig:summary}
\end{figure}

The median word length of all words is 4 characters (Fig.~\ref{fig:summary}A) for all years from 2009 to 2012. Although the median length of the 1000 most frequently used words from 2009 to 2012 is constant at 4 characters (Fig.~\ref{fig:summary}B), the peak (mode) moved from 4 characters in 2009 to 3 characters (Fig.~\ref{fig:summary}C) in the succeeding years. Based on Kruskal-Wallis tests, the word length distributions of the 1000 most frequently used words for 2010--2012 are not significantly different ($H=1.112$, $p=0.5734$) with each other but are significantly different with the distribution for 2009 ($H=10.31$, $p=0.0161$). However, the observed shortening is not just due to a sudden shortening of the 1000 most frequently occurring words from 2009 to 2010 because it was still observed in 2010--2012 (Table~\ref{tab:slopes} and Fig.~\ref{fig:temporal-summary}, 2010--2012)

From $60.32 \pm 0.0185$\% in 2009, the relative occurrence of the 1000 most frequently used words (Fig.~\ref{fig:summary}D) with respect to all words decreased to  $52.80 \pm 0.0267$\% in 2012. In that same timespan, the median utterance length in words decreased from 8 words to 5 words (Fig.~\ref{fig:summary}E) while the median tweet length in words (Fig.~\ref{fig:summary}F) decreased from 10 words to 8 words.

A topic is a word, usually in the form of \texttt{\#topic}, or a phrase that is contained in a tweet. Trending topics are the most prominent topics being talked about in Twitter within a period of time. The shortening of trending topics could potentially explain the observed shortening of utterances but instead of decreasing, the median length of trending topics increased from 11 characters in 2009 to 13 characters in 2012 (Fig.~\ref{fig:summary}G). Utterances about a trending topic are shortening but the $r^2$-values (Table~\ref{tab:slopes} and Fig.~\ref{fig:temporal-summary}, trending topics) are too small to cause the observed shortening of utterances.

The shortening of utterances is a global phenomenon and is not restricted to the US since utterances that were geolocated outside the US also exhibited shortening (Table~\ref{tab:slopes} and Fig.~\ref{fig:temporal-summary}, outside US). It was previously observed in utterances from movies and books~\cite{alis_adaptation_2012} albeit at a rate 1 and 3 orders of magnitude smaller (-0.266 char./year in books; -0.001897 char./year in movies), respectively.  Although conversations do tend to get shorter in time, our current findings show that it is occurring faster now on Twitter.

\subsection*{Geographical variation of utterance lengths}

Out of the 229 million utterances, only 795,048 utterances (0.347\%) have geographic information pointing to one of the US states (\texttt{utterances-byloc.txt} in \texttt{SI}). The number of geolocated utterances per US state is strongly correlated ($r^2=0.944$) with the 2010 census population of the US state and ranges from 396 utterances in Wyoming to 96,120 utterances in California. The medians are not correlated with the number of utterances ($r^2=0.104$) although resampling to 300 utterances, a slightly smaller number of tweets than the smallest sample size, resulted to changes in the quartiles, unlike in the previous section where resampling did not change the quartiles for almost all days after resampling. 

\begin{figure}[!ht]
\begin{center}
\includegraphics{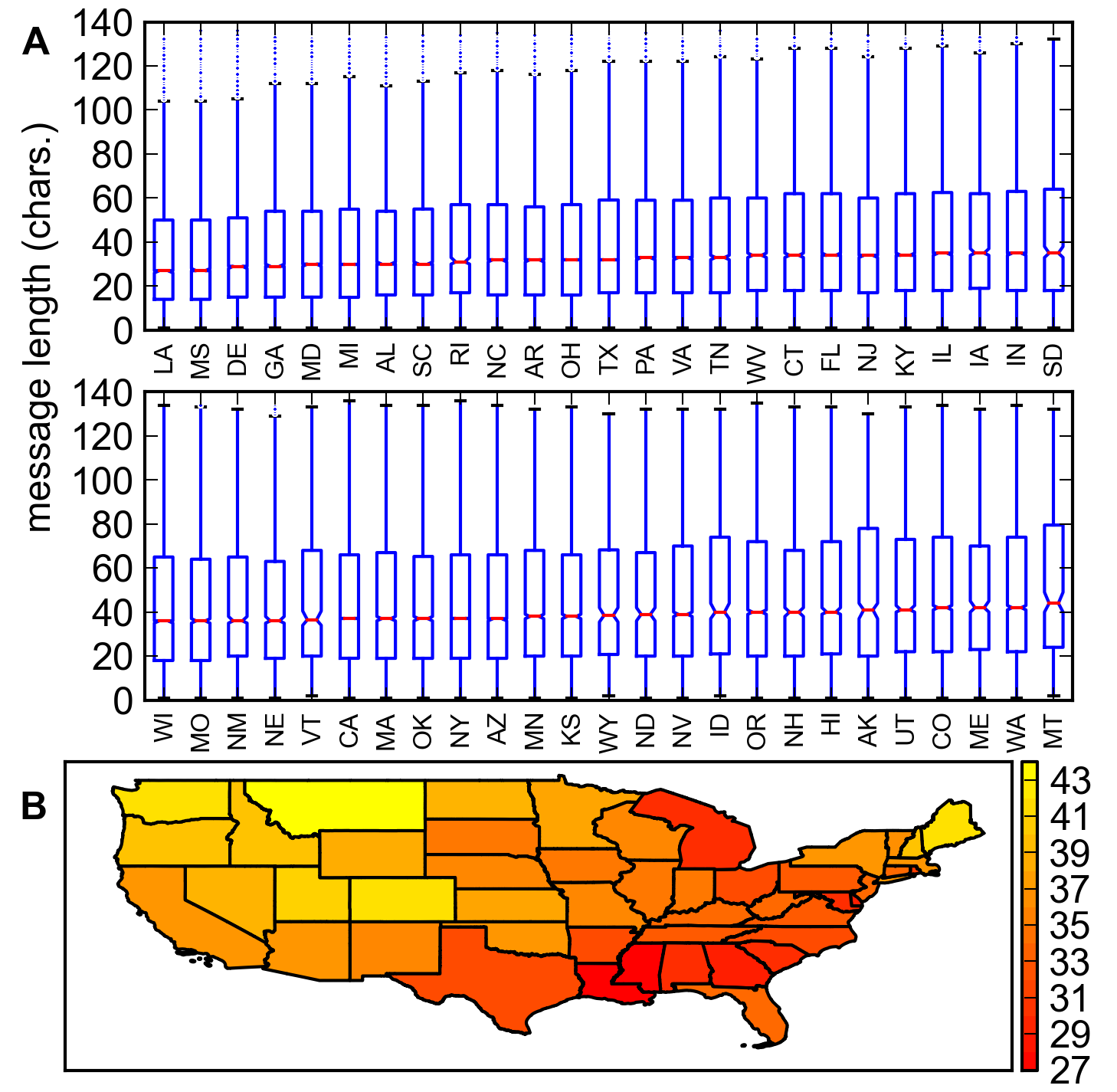}
\end{center}
\caption{
{\bf Utterance lengths across US states.}  \textbf{A.} Box plot of the utterance length distribution of each US state sorted by increasing median utterance length. The notches were estimated using 10,000 bootstrap repetitions but the resulting bootstrapped median values are the same as the empirical median values \textbf{B.} Contiguous US states colored with the bootstrapped median utterance length.
}
\label{fig:geoloc-boxplot}
\end{figure}

To estimate how the quartiles change, the quartiles were bootstrapped using $10^4$ repetitions but the bootstrapped values (Fig.~\ref{fig:geoloc-boxplot}A) turned out to be the same as the empirical values. The spread in the bootstrapped medians is very small that the interquartile range (IQR=Q3-Q1) of 40\% of the bootstrapped medians is zero. Any difference, therefore, in the median between two US states is almost guaranteed to be significant. Both Kruskal-Wallis H-test ($H=8011$,  $p<10^{-3}$)~\cite{kruskal_use_1952} and pairwise Mann-Whitney U-test~\cite{mann_test_1947} on the empirical ULD of each US state conclude that not all ULD of the US states are the same.

Plotting the medians over a US map (Fig.~\ref{fig:geoloc-boxplot}B) suggests southeastern and eastern US states tend to have shorter utterance lengths. This clustering of neighboring US states is very tenous, however, since pairwise Mann-Whitney U-tests on the median utterance length of each US state yielded non-neighboring US state pairings.

\begin{figure}[!ht]
\begin{center}
\includegraphics{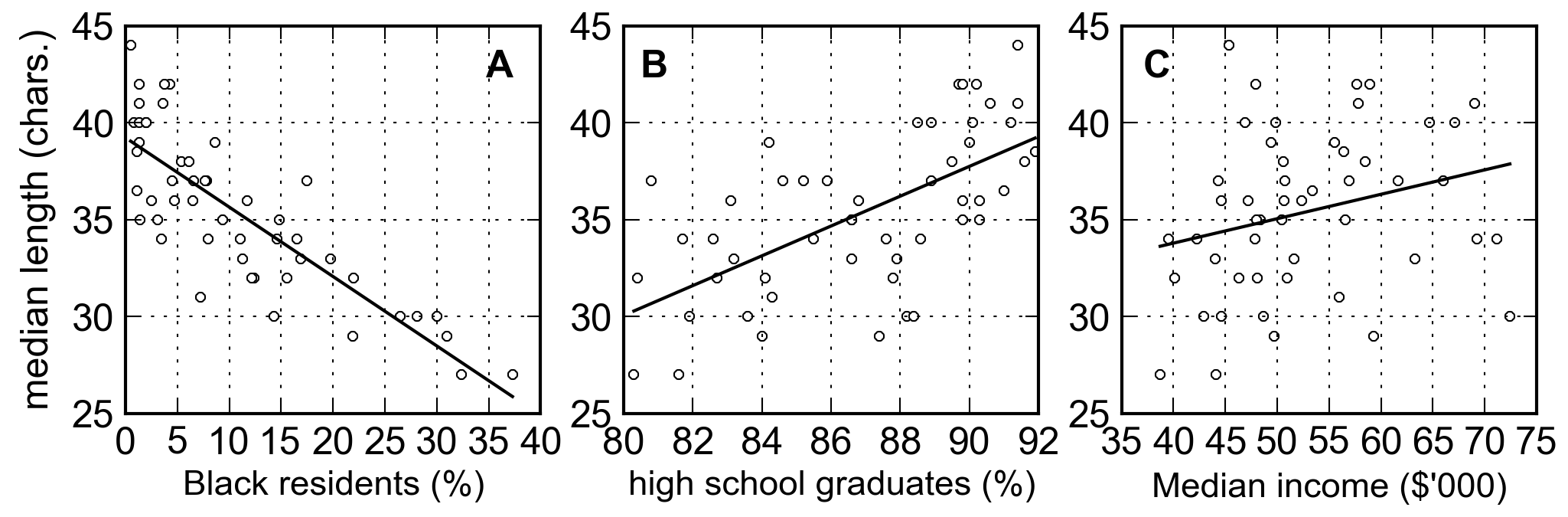}
\end{center}
\caption{
{\bf Median utterance length against demographic and socioeconomic variables.} The bootstrapped median utterance length plotted against \textbf{A.} 2011 resident Black population in percent ($r^2=0.685$), \textbf{B.} persons 25 years and over who are high school graduates or higher from 2007 to 2011, in percent ($r^2=0.397$) and \textbf{C.} Median household income from 2007 to 2011 in thousands of dollars ($r^2=0.068$). The linear regression line is also shown in each plot.
}
\label{fig:geoloc}
\end{figure}

To check for possible correlates, the bootstrapped median utterance length was regressed with demographic and socioeconomic information available in the United States Census Bureau State and Country QuickFacts~\cite{u.s._census_bureau_state_2013} (Table~\ref{tab:regression}). Out of the 51 variables (listed in SI Text S1), only the percent  Black resident population (latest data from 2011, $r^2=0.685$) and percent Black-owned firms (latest data from 2007, $r^2=0.613$) have $r^2>0.5$. A detailed description of both variables are in SI Text S1. The two variables are strongly correlated though ($r^2=0.947$) so the correlation of the bootstrapped median is really with the percentage distribution of Black residents. The bootstrapped median is inversely proportional to the Black resident population (Fig.~\ref{fig:geoloc}A). Restricting utterances to English and removing URLs improved the correlation to $r^2=0.707$.

For comparison, the median utterance length was also plotted against the percent of persons 25 years and over who are high school graduates or higher from 2007 to 2011 (Fig.~\ref{fig:geoloc}B) and median household income from 2007 to 2011 in thousands of dollars (Fig.~\ref{fig:geoloc}C), which are both described in detail in SI Text S1. Both variables are uncorrelated or only slightly correlated, at best, with the median utterance length because the values of the coefficient of determination are $r^2=0.397$ and $r^2=0.068$, respectively.

\subsection*{Multivariate regression of median utterance length}

We explored the possible dependence of the median utterance length on several variables by considering linear combinations of the QuickFacts variables. To ease the comparison of variable effect size and to avoid numerical problems, the variables were standardized by subtracting the sample mean for the variable then dividing by the sample standard deviation for the variable. That is, the $z$-scores of the variables were considered in the multiple regression. Aside from standardization, no other transformation e.g., power transformation, was performed on the variables.

The parameter estimates of the linear model (Model 2) with percent Black resident population $B$, percent high school graduates $H$, and median household income $I$ as predictors are shown in Table~\ref{tab:multiregression}. Only the coefficient for $B$ is significantly different from zero and its magnitude is 5 to 10 times larger than the coefficients for $H$ and $I$. Further supported by an F-test ($F=2.87$, df = 2, $p=0.067$, $\alpha=0.05$), the three-variable model can be simplified into the one-variable model.

Performing a stepwise regression ($\alpha_\text{in} = \alpha_\text{out} = 0.05$) with race, educational attainment and income QuickFacts variables as candidate predictors will yield a two-variable model, Model 3. The variable $B$ is still included in the model and the magnitude of its coefficient (-3.30) is about 4.5 times that of the other predictor (0.73), percent of persons 25 years and over who are holders of bachelor's degree or higher from 2007 to 2011 (denoted as variable $C$). The $r^2$ value using only $C$ as predictor is 0.093. The two-variable model cannot be reduced to a single-variable model with either $B$ ($F=5.01$, df = 2, $p=0.030$, $\alpha=0.05$) or $C$ ($F=2.87$, df = 2, $p=0.067$, $\alpha=0.05$) as the only predictor. The two-variable model improved $r^2$ by 0.03 or 4.3\% from that of the single-variable model with $B$ as the only predictor.

Expanding the set of candidate variables to all QuickFacts variables then performing another stepwise regression ($\alpha_\text{in} = \alpha_\text{out} = 0.05$) results to a five-variable model, Model 4. Both variables $B$ and $C$ are included in the model with $B$ still having the largest coefficient magnitude. By adding three more predictors to Model 3, $r^2$  increased by 0.121 or 16.9\%. The adjusted $r^2$ of Model 4 is larger by 0.114 or 16.2\% than the two-variable model and, since the latter is not equivalent to the former ($F=10.8$, df = 3, $p<10^{-3}$, $\alpha=0.05$), the former can be considered as a better model despite having more predictors. Model 4 suggests that shorter utterances are correlated with US states having larger percentage of Blacks and lower percentage of bachelor's degree holders but has more owner-owned houses, larger manufacturing output and less dense population.

The values of QuickFacts variables are regularly updated by the US Census Bureau and only the most recent values are retained. By looking up each variable in the source dataset, one can reconstruct the QuickFacts for previous years (up to 2010). Repeating the regression analysis for the different models using the data for previous years resulted to coefficient estimates that are within the standard errors of the quoted variables above. Thus, the coefficients remained essentially the same from 2010 to 2012.

\section*{Discussion}

The observation of geographic variability is not entirely unexpected
because of the existence of dialects. What is more surprising is that
the utterance length is (anti)correlated with the resident Black
population. This factor also dominates other  predictors when combined
with other demographic and socioeconomic factors using multiple
regression. 
A possible explanation is that Blacks converse more
distinctly and more characteristically than other racial groups. Since
utterances are only weakly correlated with median income and
educational attainment then perhaps the shorter utterance lengths is a
characteristic of their race---perhaps pointing towards the
controversial language of Ebonics~\cite{blommaert_ebonics_1999}. The strong correlation does 
not imply causality, and it is beyond the scope of this work to look for actual evidence of Ebonics in the tweets.

Results show that people are communicating with fewer and shorter words. The principles of least effort communications~\cite{cancho_least_2003} provide us with two possible implications. If the information content of each word remains the same then the information content of each utterance is lesser and more utterances are needed to deliver the same amount of information---a phenomenon that could be verified by tracking the complete conversations between individuals, and not just samples as we are doing now. On the other hand, if the amount of information content of each utterance remains the same then encoding becomes either more efficient (comprehension remains the same) or more ambiguous through time. When ambiguity increases,  speaker effort is minimized at the expense of listener effort. 

Based on anecdotal evidence, replies broken into several tweets are not more frequent than before but shorter spelling and omission of words do seem to be more prevalent. That is, encoding appears to become more efficient without sacrificing as much precision.

The shortening, it seems, can be explained by increased usage of jargon, which in turn provides evidence of segregation into groups. People who are engaged in a conversation communicate using a shared context, which may utilize a more specialized lexicon (jargon or even
coined words). Although utterances are expected to be less clear due to the use of fewer words, the use of context prevents this from happening. The decrease in the frequency of words from 60.32\% to 52.8\% could mean that the use of jargon increased by about 60.32\% - 52.8\% = 7.52\%. Furthermore, one of these groups might be composed of African Americans hence the dependence on percent Black population can be readily explained. Since no other demographic or socioeconomic variable is correlated with utterance length then these groupings cannot be entirely demographic or socioeconomic in nature. 

There is no obvious remaining factor that could bias the temporal analysis of utterance lengths after the shortening was shown to be robust. There are several approaches in determining the proper location of users from tweets~\cite{eisenstein_latent_2010,mocanu_twitter_2013} but we used the simplest method of assigning the location of the user to the location of the tweet. The geolocated tweets are relatively few and the tweets (users) were then aggregated by US state. Statistical data from the US Census Bureau were then used in the analysis. The inherent assumption, therefore, is that the sample used by the Bureau can also be used to describe the sample of Twitter users.  A survey done by Smith and Brenner~\cite{smith_twitter_2012}, however, showed that among the different races, Blacks significantly use Twitter more than other races. This could be the reason why only the dependence on Black population was observed. More data are needed to verify if our assumption is justified but our results are tantalizing enough to warrant a second look.

\section*{Materials and Methods}

Tweets were first retrieved using the Twitter streaming application programming interface~\cite{kalucki_streaming_2010} and corresponds to 15\% (before August 2010), 10\% (between August 2010 to mid-2012) or 1\% (mid-2012 to present) of the total public tweets. For ease of computation, we analyzed only tweets posted every Friday from 18 September 2009 to 14 December 2012. Although issues in our retrieval process prevented us from getting the entire sampled feed for the entire data collection period, only Fridays with uninterrupted and complete data were considered resulting to a total of 124 days analyzed. Conversational utterances in the form of replies were selected by filtering for tweets that begin with an at sign (@), which yielded 229 million utterances (\texttt{utterances-bydate.txt} in \texttt{SI}).

The utterance length of a tweet was measured by first stripping off all leading @usernames with the python regular expression \texttt{((\textasciicircum|\textbackslash s)*@\textbackslash w+?\textbackslash b)+}. The utterance length is the number of remaining characters after leading and trailing whitespace characters were removed. Utterances having lengths equal to zero (0.483\%) and greater than the maximum length of 140 char. (0.0935\%) were excluded from the dataset.

Tweet language identification was performed using \texttt{langid.py}~\cite{lui_langid._2012}, which claims 88.6\% to 94.1\% accuracy when identifying the language of a tweet over 97 languages. \texttt{ldig}~\cite{nakatani_short_2012} stands to be the most accurate automated language identification system for tweets having a claim of 99\% accuracy over 19 languages ($>98\%$ accuracy for English), however, it has not yet been formally subjected to peer review. Nevertheless, we repeated the analysis using \texttt{ldig} and found similar results and conclusions.

Tweets were geolocated using the \texttt{geo} and \texttt{coordinates} metadata of the tweets and were categorized by US states using TIGER/Line shapefiles~\cite{us_census_bureau_2012_2012} prepared by the US Census Bureau. A user must opt-in to have location information be attached to their tweets. Previously, only exact coordinates (latitude and longitude) are attached as location information and these become the values of the \texttt{geo} and \texttt{coordinates} metadata of the tweet. More recently, users may opt to select less granular location information e.g., neighborhood, city and country, and these less precise place information are now the default.~\cite{twitter_faqs_2013} A user, though, may still choose exact location information or omit location information for every tweet. Geolocation is possible with both mobile clients and browsers but geolocation for the latter is not yet available for all countries.

A parallel~\cite{cafaro_finding_2011} version of the \textit{Space Saving}~\cite{metwally_efficient_2005} algorithm for selecting the most frequent $k$ words was used instead of a naive histogram of word occurrences because of the prohibitive amount of resources needed. The \textit{Space Saving} algorithm maintains a frequency count of up to $k$ words only. An untracked word replaces the least frequent word if the maximum number of $k$ words are already being tracked. The parallel \textit{Space Saving} algorithm involves partitioning the data then running the \textit{Space Saving} algorithm for each chunk. The results of each chunk are merged using an algorithm similar to \textit{Space Saving}. The word frequency of both \textit{Space Saving} and its parallel version are approximate for near-$k$-ranked words. To have a guaranteed list of the 1000 most frequently occurring words, a much larger value of $k=10^5$ was used.

\section*{Acknowledgments}

Some computational resources were provided by an AWS in education grant and the Advanced Science and Technology Institute, Department of Science and Technology, Philippines.


\section*{Figure Legends}

\section*{Supporting Information Legends}

Text S1. Information on State and County QuickFacts variables.

Dataset S1. Utterance length frequencies by date. The rows of this comma-separated file correspond to tweets posted on a certain UTC date. The first column is the date in ISO format (yyyy-mm-dd) and the remaining columns list the number of utterances with a length of 1 character, 2 characters, 3 characters and so on, until 139 characters. Only the frequencies of ``valid" utterance lengths (1-139 characters) are included.

Dataset S2. Utterance length frequencies by US state. The rows of this comma-separated file correspond to tweets posted from a US state. The first column is the abbreviated US state (e.g., AK) and the remaining columns list the number of utterances with a length of 1 character, 2 characters, 3 characters and so on, until 139 characters. Only the frequencies of ``valid" utterance lengths (1-139 characters) are included.

\section*{Tables}

\begin{table}[!ht]
\caption{
\bf{Slopes of utterance length quartiles temporal regression lines}}
\begin{tabular}{|l|*{6}{c|}}
Subset & \multicolumn{2}{|c|}{Q1} & \multicolumn{2}{|c|}{Median (Q2)} & \multicolumn{2}{|c|}{Q3} \\
& Slope & $r^2$ & Slope & $r^2$ & Slope & $r^2$ \\
& (chars./year) & & (chars./year) & & (chars./year) & \\
All & -2.53 & 0.916 & -5.20 & 0.926 & -8.32 & 0.862 \\
Sep--Dec & -5.29 & 0.812 & -7.85 & 0.894 & -9.97 & 0.889 \\
Resampled & -2.54 & 0.918 & -5.25 & 0.927 & -8.23 & 0.860 \\
URLs removed & -2.42 & 0.933 & -4.63 & 0.922 & -7.51 & 0.887 \\
English only & -3.38 & 0.910 & -5.95 & 0.881 & -8.35 & 0.785 \\
2010--2012 & -5.19 & 0.842 & -8.07 & 0.910 & -10.4 & 0.938 \\
Trending topics & -3.41 & 0.153 & -6.83 & 0.294 & -7.61 & 0.440 \\
Outside US & -5.57 & 0.838 & -8.15 & 0.904 & -10.4 & 0.909\\
\end{tabular}
\label{tab:slopes}
 \end{table}
 
\begin{table}[!ht]
\caption{
\bf{Single-variable linear regression of median utterance length with selected US Census Bureau QuickFacts variables}}
\footnotesize
\begin{tabular}{|p{0.2\textwidth} *7{|c}|}
\textbf{Independent variable} & \multicolumn{7}{|c|}{\textbf{Parameter estimate (standard error)}} \\
 & \textbf{1a} & \textbf{1b} & \textbf{1c} & \textbf{1d} & \textbf{1e} & \textbf{1f} & \textbf{1g} \\
 
2011 resident Black population in percent $B$ & \textbf{-3.411***} & & & & & & \\
 & \textbf{(0.334)} & & & & & & \\ 
 
Persons 25 years and over who are high school graduates or higher from 2007 to 2011 in percent $H$ & & \textbf{2.597***} & & & & & \\ 
 & & \textbf{(0.462)} & & & & & \\ 
 
Median household income from 2007 to 2011 in thousands of dollars $I$ & & & 1.074 & & & & \\
 & & & (0.574) & & & & \\

Persons 25 years and over who has bachelor's degree or higher from 2007 to 2011 in percent $C$ & & & & \textbf{1.254**} & & & \\
 & & & & \textbf{(0.567)} & & & \\

2010 population per square mile $D$ & & & & & \textbf{-1.247**} & & \\
 & & & & & \textbf{(0.567)} & & \\
 
Owner-occupied housing units in percent of total occupied housing units from  2007 to 2011 $O$ & & & & & & -0.846 & \\
 & & & & & & (0.582) & \\
 
Total value of manufacturing shipments in 2007 $M$ & & & & & & & \textbf{-1.310**} \\
 & & & & & & & \textbf{(0.564)} \\

Constant & \textbf{35.40***} & \textbf{35.40***} & \textbf{35.40***} & \textbf{35.40***} & \textbf{35.40***} & \textbf{35.40***} & \textbf{35.40***} \\
 & \textbf{(0.334)} & \textbf{(0.462)} & \textbf{(0.574)} & \textbf{(0.567)} & \textbf{(0.567)} & \textbf{(0.582)} & \textbf{(0.564)} \\

$r^2$ & 0.685 & 0.397 & 0.068 & 0.093 & 0.092 & 0.042 & 0.101 \\

adjusted $r^2$ & 0.678 & 0.384 & 0.049 & 0.074 & 0.073 & 0.022 & 0.082 \end{tabular}
\begin{flushleft}
Standard errors are presented in parentheses below the corresponding parameter estimates. Bold indicates significance at the 5\% level, $n=50$\\
** $p < 0.05$\\
*** $p < 0.01$
\end{flushleft}
\label{tab:regression}
 \end{table}

\end{document}